\documentclass[conference]{IEEEtran}

\IEEEoverridecommandlockouts
\usepackage{amsmath,amssymb,amsfonts}
\usepackage{cite}
\usepackage{algorithmic}
\usepackage{graphicx}
\usepackage{textcomp}
\usepackage{xcolor}
\usepackage{svg}
\usepackage{footmisc}

\def\BibTeX{{\rm B\kern-.05em{\sc i\kern-.025em b}\kern-.08em
    T\kern-.1667em\lower.7ex\hbox{E}\kern-.125emX}}

\begin{document}

\title{Détection multidomaine d'anomalies\\ dans un réseau 5G\\
%{\footnotesize \textsuperscript{*}Note: Sub-titles are not captured in Xplore and
%should not be used}

%\thanks{Identify applicable funding agency here. If none, delete this.}
}

\author{\IEEEauthorblockN{Thomas Hoger}
\IEEEauthorblockA{\textit{LAAS - CNRS, Université de Toulouse, CNRS} \\
Toulouse, France \\
thomas.hoger@laas.fr}
\and
\IEEEauthorblockN{Philippe Owezarski}
\IEEEauthorblockA{\textit{LAAS - CNRS, Université de Toulouse, CNRS} \\
Toulouse, France \\
owe@laas.fr}
% \and
% \IEEEauthorblockN{3\textsuperscript{rd} Given Name Surname}
% \IEEEauthorblockA{\textit{dept. name of organization (of Aff.)} \\
% \textit{name of organization (of Aff.)}\\
% City, Country \\
% email address or ORCID}
% \and
% \IEEEauthorblockN{4\textsuperscript{th} Given Name Surname}
% \IEEEauthorblockA{\textit{dept. name of organization (of Aff.)} \\
% \textit{name of organization (of Aff.)}\\
% City, Country \\
% email address or ORCID}
% \and
% \IEEEauthorblockN{5\textsuperscript{th} Given Name Surname}
% \IEEEauthorblockA{\textit{dept. name of organization (of Aff.)} \\
% \textit{name of organization (of Aff.)}\\
% City, Country \\
% email address or ORCID}
% \and
% \IEEEauthorblockN{6\textsuperscript{th} Given Name Surname}
% \IEEEauthorblockA{\textit{dept. name of organization (of Aff.)} \\
% \textit{name of organization (of Aff.)}\\
% City, Country \\
% email address or ORCID}
}

\maketitle

\begin{abstract}
Avec l’avènement de la 5G, les réseaux mobiles se dynamisent et vont donc présenter une plus large surface d'attaque. Pour sécuriser ces nouveaux systèmes, nous proposons une méthode de détection d'anomalies multidomaine qui se distingue par l'étude de la corrélation du trafic sur trois dimensions : temporel en analysant les séquences de messages, sémantique en abstrayant les paramètres que ces messages contiennent et topologique en les reliant sous forme de graphe. Contrairement aux approches traditionnelles qui se limitent à considérer ces domaines indépendamment, notre méthode étudie leurs corrélations pour obtenir une vision globale, cohérente et explicable des anomalies. 

%Nous visons également à créer un dataset intégrant des données de trafic issues de plusieurs niveaux du réseau afin de permettre une analyse approfondie multidomaine.

%Les changements de la 5G augmentent drastiquement la surface d'attaque des réseaux du futur et introduisent de nouvelles vulnérabilités qui leur sont propres. 
\end{abstract}


\section{Introduction}
Les réseaux mobiles sont historiquement conçus comme des systèmes monolithiques où chaque fonction est exécutée sur un matériel dédié, communiquant avec ses intermédiaires par des canaux fixes et des piles protocolaires figées. Ces limites disparaissent en 5G avec la virtualisation des fonctions réseaux (NFV) qui permet de déployer des services à la demande sur n’importe quel type de matériel. La structure du réseau de cœur peut être définie logiciellement et la 5G passe alors à un système de client-serveur qui suit le principe REST. Les procédures permettant le bon fonctionnement du réseau deviennent donc des microservices déployables dynamiquement et n’importe quelle entité peut ensuite les solliciter par le biais de protocoles de haut niveau, souvent HTTP. 
%% Nouveau
Néanmoins l’aspect dynamique des services de la 5G et la complexification de ses protocoles augmentent grandement sa surface d’attaque. Ainsi, bien que la 5G partage les vulnérabilités d'un réseau classique, elle est maintenant aussi la cible de nouvelles attaques qui lui sont propres. Coldwell et al. ont notamment trouvé des vulnérabilités sur les paramètres des appels d'API\cite{5gad} en profitant du manque d'authentification des fonctions dans le réseau de cœur. Amponis et al. mettent en lumière des attaques semblables sur PFCP, un protocole dédié à la 5G \cite{pfcp}. Garbelini et al. proposent quant à eux des attaques de déni de service sur des protocoles bas niveau \cite{5ghoul}. Comme la détection de ces attaques nécessite une inspection détaillée des paquets, une analyse quantitative de flux comme celles réalisées par les IDS classiques ne suffit plus. Pour sécuriser la 5G, il nous faut donc concevoir des méthodes de détection qui sauront s’adapter à un système dynamique. 
%% Partie ML
Pour détecter une anomalie dans un tel système, un humain s’appuie sur plusieurs éléments clés comme les informations sur un paramètre, aussi bien au niveau de son nom que de sa valeur. L'analyse des paramètres et de leur sémantique permet de remarquer les paquets avec des valeurs aberrantes ou impossibles. Cependant, il arrive que le contenu même du paramètre soit moins pertinent que son contexte comme dans le cas des adresses IP, hash ou divers identifiants. Il existe ici deux types de contextes que l'on peut observer : le contexte temporel et le contexte topologique. Le contexte temporel analyse les séquences pour révéler certaines anomalies, tel que les DoS par rejeu de messages légitimes. Le contexte topologique, quant à lui, observe les liens entre les différents acteurs, ce qui permet de détecter des attaques coordonnées comme des DDoS. Notre objectif est de nous inspirer de la façon dont un humain détecterait une anomalie, en intégrant trois dimensions à notre détection : la sémantique du contenu, le contexte topologique et le contexte temporel. 

%Ces méthodes devront permettre une analyse multi-dimensionnelle des données, à la fois sémantique à travers les valeurs des paramètres, mais aussi topologique avec les interactions entre les différents acteurs puis enfin temporel avec l'ordre d'arrivée des messages.
%\section{Cadre d'étude}

% ------------------------------------------

\section{Cadre de l'étude}
Le design dynamique de la 5G a pour but de permettre aux opérateurs réseaux de déployer de nouveaux services rapidement. Cela permet entre autres d'incorporer à son réseau des fonctions personnalisées correspondant chacune à des besoins spécifiques. Le nombre de cas d'utilisations et de fonctions associées ne va cesser de croître et il est donc nécessaire de définir des limites claires pour le cadre de notre étude. Pour cela, nous choisissons de nous intéresser seulement aux éléments implémentés par le projet Open Air Interface (OAI)\cite{oai_cn} que nous utiliserons pour nos expérimentations. 

\subsection{Scénarios d'intrusion}
Les attaques de l'état de l'art nécessitent en grande majorité d'avoir au préalable mis en place un point d'entrée dans le réseau 5G, que ce soit en compromettant une machine existante ou en intégrant un nouvel acteur hostile. Il existe cependant de nombreuses méthodes pour arriver à remplir ces conditions. Un attaquant peut par exemple profiter de vulnérabilités sur les slices\cite{slice2} pour pivoter sur des machines d'un réseau auquel il n'avait auparavant pas accès. Comme les fonctions réseaux sont virtualisées, il est aussi possible de les atteindre via leur hôte\cite{docker}. Un attaquant pourrait, pour finir, profiter de la proximité entre les stations de base 5G (gNB) avec les utilisateurs pour accéder physiquement à l'une d'entre elle et en prendre le contrôle. Nous faisons donc l'hypothèse préliminaire qu'un attaquant a réussi à s'introduire sur le réseau en prenant le contrôle d'un ou plusieurs équipements utilisateurs (UE), gNB ou fonctions du réseau de cœur (CN).

\subsection{Chiffrement des messages}
Pour faciliter notre étude nous adoptons le point de vu de l'opérateur administrant le réseau 5G et considérons que le trafic de contrôle qu'il intercepte est lisible en clair. En effet, même si la communication entre NFV est chiffré de point à point par le protocole TLS, l'opérateur a le contrôle de chaque machines du CN et peut donc déchiffrer les messages à sa guise. L'attaquant n'ayant quant à lui pas le contrôle du réseau, on considère qu'il n'est pas capable de déchiffrer le trafic récupéré par écoute passive. Nous limitons donc notre étude aux attaques nécessitant une participation active d'un attaquant. 

\subsection{Surfaces considérées}
Nos analyses concernent aussi bien la couche applicative que des protocoles de bas niveaux.  
% Applicative
Au niveau applicatif, il est seulement possible de traiter les données utilisateur avec de la volumétrie. En effet, une fois la connexion avec l'UE établie, le réseau 5G sert de tunnel au trafic utilisateur, ce qui rend impossible de déchiffrer le contenu des échanges et donc de l'analyser en profondeur. Comme l'analyse de volumétrie est déjà largement prise en compte par les systèmes de détection d'intrusion (IDS) classiques, nous préférons consacrer notre étude de la couche applicative au trafic de contrôle dont le rôle est essentiel au bon fonctionnement de l'intégralité du réseau. Parmi les fonctions réseaux agissant sur le flux de contrôle nous choisissons de nous limiter dans un premier temps aux AMF, AUSF, NRF, SMF, UDM, PCF, UDR et UPF\footnote{Ces fonctions réseau sont respectivement responsables de la gestion de l'accès et de la mobilité, de l'authentification, de la gestion des informations du réseau, de la gestion des sessions, de la gestion des données d'abonnés, du contrôle des politiques, du stockage des données d'abonnés et du traitement du plan utilisateur.}.
%AMF : Access and Mobility Management Function, AUSF : Authentication Server Function, NRF : Network Repository Function, SMF : Session Management Function, UDM : Unified Data Management, PCF : Policy Control Function, UDR : Unified Data Repository, UPF : User Plane Function
Nous considérons que ces fonctions sont les éléments fondamentaux d'un réseau 5G et seront retrouvées dans toutes les implémentations quel que soit l'opérateur qui la met en place. Pour chacune, nous considérons l'ensemble des procédures supportées par OAI et détaillées dans leur documentation respective.
% Réseau / Liason
Pour les niveaux réseau et liaison nous nous concentrons là aussi sur un sous ensemble des protocoles, ou versions de protocoles, spécifiques à la 5G. Parmi eux on retrouve notamment NGAP, NAS, RRC et RLC. 

%Vecteur d'entrée (la plupart des attaques qu'on considère nécessittent l'intrusion préalable dans le réseau mais c'est possible de tel façon)
%Chiffrement (on se positionne comme le boss au milieu donc on peut déchiffrer le point à point)
%Limites des protocoles considérés (que ceux qui sont supportés par l'implem oai)

% ------------------------------------------

\section{Approche proposée}
\begin{figure*}[!tbp]
\centerline{\includegraphics[width=0.8\linewidth]{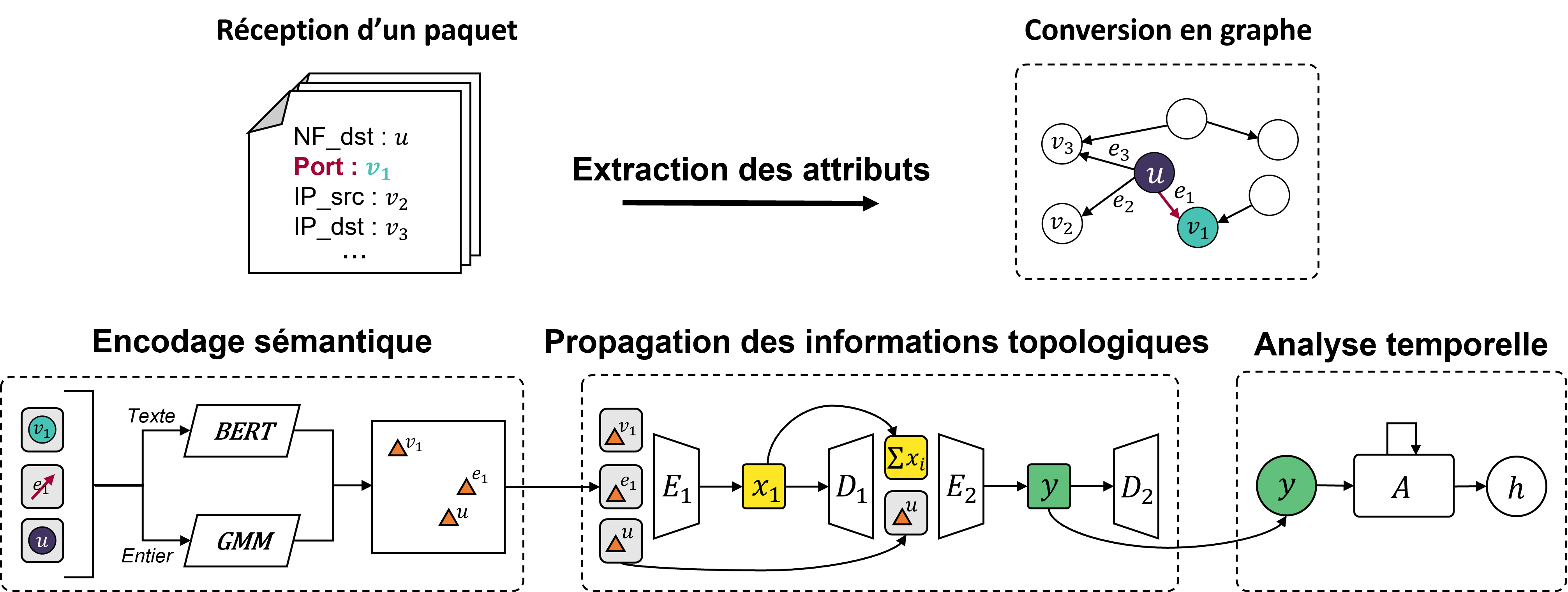}}
\caption{Pipeline pour la détection d'anomalies multidomaines. Chaque paquet est ajouté dans un graphe en tant que nœud central \(u\) avec comme valeur le nom de l'entité réceptrice. On relie ensuite ce nœud central \(u\) à des nœuds successeurs \(v_{i}\) représentant chacun un attribut que le paquet contient. L'arc \(e_{i}\) contient le nom de l'attribut (par exemple ``ip``) tandis que le nœud successeur \(v_{i}\) contient la valeur de l'attribut (par exemple ``192.168.1.1``). Le contenu de ces différents nœuds et arcs est projeté dans un espace vectoriel par un encodeur dont la nature dépend du type de valeur rencontré (NLP pour le texte, GMM pour les entiers). 
%Pour tous les successeurs du nœud central on transmet un triplet d'attributs encodés \((u_i,e_{ij},v_{ij})\) à un auto encodeur qui condensera ces informations en un vecteur représentant des parties du voisinage. 
Pour chaque successeur du nœud central, un triplet d'attributs encodés est transmis à un auto encodeur composé de \(E_{1}\) et \(D_{1}\) sous la forme \((u,e_{i},v_{i})\). Celui-ci se charge de condenser le triplet en un vecteur \(x_{i}\) unique, chacun représentant une partie spécifique du voisinage du nœud central. Ces condensats sont ensuite agrégés en un vecteur \(\sum x_{i}\) et ré-encodés avec le nœud \(u\). Le second auto encodeur composé de \(E_{2}\) et \(D_{2}\) produit ensuite une représentation \(y\) globale du nœud et de son voisinage. On donne enfin cette représentation à un RNN \(A\) qui analysera les séquences de messages.}
% On propage ensuite les informations topologique grâce à des auto-encodeurs auxquels on fourni un triplet de valeurs encodées
\label{fig}
\end{figure*}

Pour détecter des anomalies sur les trois domaines que nous considérons nous avons choisi de combiner successivement des modèles spécialisés. Nous avons préféré une analyse successive à une analyse parallèle car les premiers modèles de notre chaîne sont des auto encodeurs qui, en plus de détecter des anomalies, permettent d'abstraire les données qu'ils traitent. Ainsi, en passant par ces modèles, les informations sont enrichies et les corrélations sont accentuées. La chaîne de traitement que nous avons conçue est présentée dans la figure \ref{fig}. 

\subsection{Analyse sémantique}
Plusieurs attaques 5G sur le CN n'ont besoin d'envoyer qu'un seul paquet pour être menées à bien. Elles ne pourront donc pas être détectées par des méthodes d'analyse quantitatives. Pour déterminer la normalité d'un paquet, il est donc fondamental d'inspecter ses paramètres et de comprendre leur nature. Cette compréhension demande de considérer conjointement la valeur et le nom des paramètres. Malheureusement, la notion de sémantique apparaît rarement dans l'état de l'art qui se concentre le plus souvent sur l'analyse des valeurs avec des modèles comme des CNN. 
% One Hot
Pour utiliser dans notre apprentissage des valeurs textuelles telles que les noms de paramètres, il est nécessaire de les encoder. Une approche possible serait d'utiliser un encodage one-hot qui transforme chaque paramètre différent en un vecteur unique et orthogonal. Cependant, il existe un grand nombre de noms de paramètres différents dans les CN 5G et certains, bien que partageant la même sémantique, présentent des variations morphologiques. C'est par exemple le cas des paramètres nfType et targetNfType qui, bien que légèrement différents, seraient intéressants à regrouper. Notre objectif est donc d’abstraire nos données textuelles en les projetant dans un espace vectoriel où leur proximité avec les autres points reflétera une similitude sémantique. 
% Word2Vec+Bert
La prise en compte de cette sémantique est possible grâce à des techniques de Natural Language Processing (NLP) tel que Word2Vec\cite{word2vec} et ses variantes. Ces méthodes apprennent la sémantique d'un mot à travers son contexte et représentent les similitudes par des proximités vectorielles. Cependant, les mots sont ici projetés indépendamment dans un espace vectoriel de même dimension que celui d'origine. En plus d'être difficilement scalable, ces méthodes peinent aussi à traiter les variations morphologiques ainsi que les nouveautés. Nous leur préférerons donc FastText\cite{fasttext}, une des variante de Word2Vec, qui découpe les mots en tokens et résout les problèmes énoncés précedemment par une analyse plus fine. Nous considérerons aussi des modèles de transformer comme BERT\cite{bert} et ses variantes, qui bien que moins rapides que FastText, compensent par une meilleure compréhension des informations contextuelles et des relations sémantiques.
% GMM
Bien que ces méthodes soient applicables aux noms et valeurs textuelles, elles sont moins efficaces lorsqu'il s'agit de valeurs numériques. Dans ce cas, nous utiliserons un Gaussian Mixture Model (GMM) qui modélise les répartitions des valeurs par des distributions gaussiennes et permet de les regrouper en plusieurs catégories. Les GMM sont particulièrement efficaces pour approximer des densités et ont déjà été utilisés avec succès dans des scénarios de détection d’anomalies \cite{sec2graph}. Pour pouvoir utiliser à la fois les résultats de GMM et du NLP, il faudra néanmoins passer par une couche ``fully connected`` qui harmonisera les dimensions de leur espace de projection.

\subsection{Analyse topologique}

%Pour cela on utilise 
% dire qu'on sort bien les attributs en arc + noeud
% dire que gnn c'est bien mais pas scalable donc on le restreint à une certaine zone

%la topologie du trafic et de représenter

\begin{figure}[!tbp]
\centering
\includegraphics[width=0.6\linewidth]{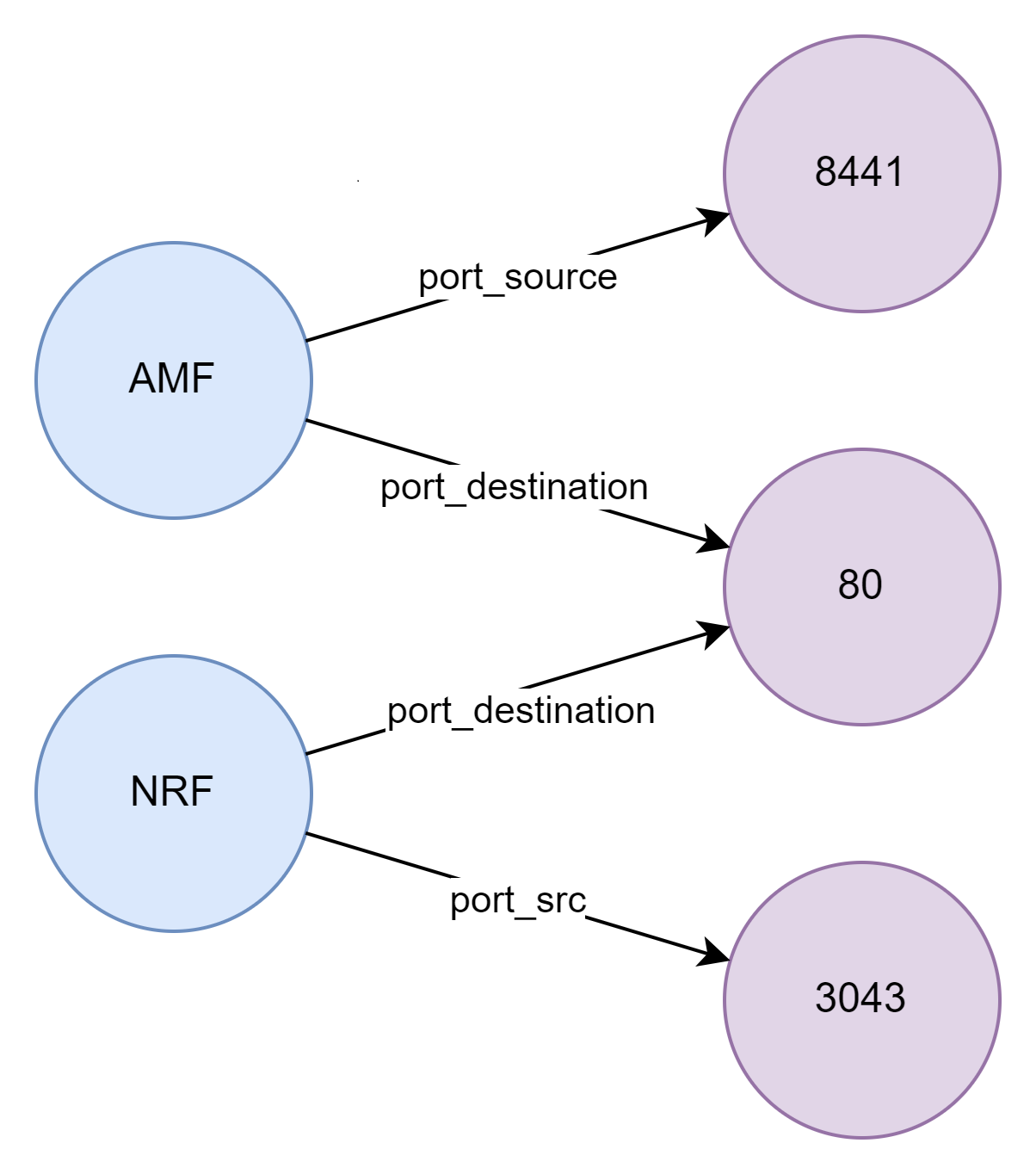}
\caption{Exemple de graphe généré par la réception de deux paquets destinés respectivement à un AMF et un NRF. Les NF, à gauche, sont liées par des arcs aux attributs que leur paquet contient. L'arc contient le nom de l'attribut tandis que sa valeur est dans le nœud cible.}
\label{graphe}
\end{figure}

Certains attributs, comme les adresses IP, les ports, les identifiants d'instance de NF ou les hachages, n'ont pas d'intérêt intrinsèque, mais doivent être considérés d'un point de vue relationnel. La forte présence d'une même IP source, quelle que soit sa valeur, pourrait par exemple être un indice d'une attaque de DDoS. Pour représenter ces relations, on peut modéliser les NF et leurs attributs par des nœuds dans un graphe. Le lien entre NF et attribut peut quant à lui être représenté par un arc. On obtient alors un graphe biparti dirigé et acyclique tel que présenté dans la figure \ref{graphe}. L'analyse topologique de ce graphe permet de comprendre la structure des échanges à grande échelle et ainsi de détecter des attaques coordonnées. Pour intégrer cette information dans la détection, nous avons décidé d'explorer les Graph Neural Networks (GNN), qui sont particulièrement adaptés pour capturer des informations topologiques. Il existe un grand nombre de techniques différentes, mais leur objectif est toujours de propager l'information sémantique d'un nœud vers ses voisins. Dans notre cas, il est crucial de prendre en compte les valeurs des arcs, ce qui nous a orienté vers les Message Passing Neural Networks (MPNN)\cite{mpnn}, des GNN intégrant explicitement ces valeurs dans leur processus. Cette technique est le plus souvent utilisée sur des graphes statiques et il nous faut donc la limiter à un sous-ensemble du graphe pour éviter une explosion combinatoire. La profondeur de propagation est donc un hyperparamètres sur lequel nous réaliserons des études expérimentales pour en déterminer une valeur adéquate.

%Une autre méthode, le PNA (Pooling-based Neural Network) \cite{pna}, offre des techniques similaires, mais au lieu d'appliquer une opération d'agrégation unique (comme min ou max), elle utilise plusieurs opérations simultanément. Cette approche améliore significativement les résultats sur les petits graphes, mais elle est moins efficace sur des graphes de grande taille. Cependant nous n'avons pas choisi la méthode PNA car elle nécessite la spécification du degré maximum des nœuds du graphe lors de la définition du modèle or ce paramètre peut varier si le graphe est dynamique. 
    
\subsection{Analyse temporelle}
Analyser les séquences de messages reçus permet de détecter les attaques de rejeu et de DoS. En effet, le trafic du plan de contrôle 5G, tel que défini par la norme 3GPP, suit des modèles d'échange séquentiels et toute déviation par rapport à ces schémas représente une anomalie. L'analyse topologique s’occupe déjà des corrélations entre les différents acteurs, ce qui nous permet de nous concentrer spécifiquement sur les comportements indépendants des acteurs. Pour ce faire, les paquets destinés à une même entité sont regroupés en séquences représentant son activité individuelle. Ces séquences sont ensuite fournies à un réseau neuronal récurrent (RNN), spécialement conçu pour analyser les corrélations temporelles.

\subsection{Explicabilité}
Si notre détection se limite à classifier les messages sans fournir d'explications, l'information perd considérablement de sa valeur. Dans le cas des remédiations automatiques, cette perte n'est pas importante mais il arrive parfois que la nature d'une anomalie soit ambiguë et nécessite l'intervention d'un spécialiste. Ce dernier devra interpréter l'anomalie et il est donc fondamental de faciliter son travail. Notre approche répond intrinsèquement à ce besoin par son design modulaire qui permet de lever des alertes sur différents niveaux du pipeline. 

% ------------------------------------------

\section{Etat de l'art}
% Sémantique 
Une première approche de détection considère les octets d'un paquet comme une image sur laquelle on utilise des réseaux de neurones convolutifs (CNN) pour détecter des variations structurelles \cite{ensemble}\cite{cnn}. L'idée de mettre à profit la puissance intrinsèque des CNN qui ont déjà eu des résultats encourageants sur des problèmes ne relevant pas directement de la computer vision, comme par exemple la détection de malwares. Cependant, cette méthode se base principalement sur la disposition des octets, sans aucune garantie que ce format soit constant à travers les implémentations de la 5G. De plus, comme les CNN n’interprètent pas le contenu des données qu'ils traitent, ils ne peuvent pas évaluer l'impact qu'aura le changement d'une valeur ni établir de lien entre paramètres similaires. 
%Ainsi, nous considérons qu'il est essentiel de développer un système capable d’interpréter la sémantique des données qu'il traite.
%Cependant, pour passer d'un paquet à une image il est nécessaire de tronquer les données ou de leur ajouter du padding, altérant ainsi les informations d'origine
%--------------
%Certaines études tentent de détecter les attaques de rejeu et de DoS en analysant les séquences de message reçu. 
% Temporel
AutoGuard\cite{autoguard} et ADSeq-5GCN\cite{adseq} utilisent quant à eux des réseaux de neurones récurrents (RNN) pour capturer les dépendances temporelles au sein de séquences de messages. Cependant ces travaux se concentrent exclusivement sur la forme des séquences et ignorent complétement le contenu des messages. 
% en plus ADSeq-5GCN met de côté le NRF alors que c'est un des modules sensible
%alors que ces informations temporelles sont complémentaires aux autres domaines d'analyse et permettent rarement de détecter des attaques à elles seules.
% Topologique
De son côté, PROV5GC \cite{prov5g} utilise des graphes pour établir des liens entre les entités communicantes. Chaque nœud est un acteur différent et les arcs qui les relient représentent des échanges de messages. Le contenu des paquets est présent dans les arcs mais il est encapsulé dans un dictionnaire JSON, ce qui empêche de lier les paramètres entre eux. GSAD\cite{gsad} fonctionne de la même manière mais considère la payload comme une suite d'octets qui sont encodés avant d'être insérés en tant que contenu des arcs de leur graphe. Dans les deux approches, la payload n'est pas interprétée et les relations établies se limitent alors aux acteurs communicants, sans prendre en compte le contenu des échanges. 
% Conclusion 
À notre connaissance, l'état de l'art ne propose aucune approche s'intéressant à la sémantique du contenu des paquets, et ne tente pas non plus de combiner simultanément les trois domaines que nous explorons.

% dans les deux cas on interprete pas la payload donc tout est ``comprimé`` et on fait des relations que sur les acteurs et pas sur les contenus. donc on passe à côté d'une grosse partie de ce qui est itnéressant

%Dans le cas de GSAD\cite{gsad}, les données de la payload sont considérés comme une séquence d'octets qui sont encodés puis insérés en tant que contenu des arcs de leur graphe. Leur formalisme partage donc les mêmes problèmes que PROV5GC et ne permet pas non plus d'établir des liens entre les paramètres. 

%Les données de la payload sont ici considérés comme une séquence d'octets qui sont traités, encodés, puis insérés en tant que contenu des arcs de leur graph. Leur méthode partage donc les problèmes les autres CNN qui ne prennent pas en compte les variation possibles de formats et de contenus dans les messages légitimes. 

\section{Perspectives}
La validation expérimentale de notre approche est une étape essentielle pour démontrer sa pertinence. Il est donc nécessaire d'avoir à disposition des données réalistes et adaptées aux besoins de notre étude. La conception d’un dataset ainsi que la génération automatique d’attaques sont donc les axes principaux que nous souhaitons explorer dans le futur.

\subsection{Datasets}
A ce jour, il existe deux principaux dataset d'attaques 5G. Le premier, 5GAD \cite{5gad} met en place des attaques sur l'API REST sur une plateforme free5GC avec deux UEs et un gNB. Malheureusement, leur version de free5C est relativement ancienne et utilise le protocole HTTP au lieu d'HTTP2. De plus, leurs données bénignes représentent uniquement du trafic utilisateur (streaming youtube, téléchargement ftp, conférence teams) alors que nous avons dans notre cas besoin de trafic de contrôle. Le deuxième ``5G Core PFCP Intrusion Detection Dataset`` \cite{pfcpdataset} à été réalisé avec Open5GC et UERANSIM. Leurs données incluent les 4 attaques qu'ils ont mises au point ainsi que des données qu'ils décrivent comme ``normales``, mais nous n'avons pas trouvé d'informations supplémentaires concernant la nature de leur trafic bénin. Notre objectif étant de détecter des attaques sur plusieurs protocoles, il nous est nécessaire d'avoir des datasets représentant chacun d'entre eux et réalisés dans un même environnement. Or il n'existe pas à notre connaissance de dataset intégrant des attaques sur plusieurs protocoles différents. Il nous est donc nécessaire de réaliser notre propre dataset de trafic de contrôle applicatif et réseau. 
% Ajouter des infos pour dire comment on va procéder en gros
% J'en suis pas super content mais bon
Pour ce faire, nous prévoyons d'utiliser la plateforme 5G du LAAS-CNRS qui met en place un réseau d'accès radio complet sur lequel est déployé le code du projet Open Air Interface \cite{oai_cn}. Cette plateforme comprend trois serveurs DELL 7920, chacun doté de 2 processeurs à 18 cœurs et d'un noyau optimisé pour la faible latence, qui servent respectivement de CN, gNB et d'UE. La plateforme inclut également trois USRP X310, chacun équipé de deux antennes log périodiques, ainsi que deux téléphones Google Pixel 6 agissant en tant qu'UE et disposant chacun de cartes SIM spéciales commercialisées par le projet Open Cells.

\subsection{Génération automatique d'attaques bas niveau}
À ce jour, les attaques sur la couche applicative nécessitent l'utilisation de paramètres précis et peuvent difficilement être automatisées. D'un autre côté, les attaques sur les couches de bas niveaux  telles que proposées par Garbelini et al.\cite{5ghoul} sont le plus souvent des injections de valeurs incorrectes dont le but est de provoquer un dysfonctionnement logiciel. Cette caractéristique se prête particulièrement bien à l’élaboration d’un modèle adverse. On pourrait avoir d'une part un générateur capable de produire des paquets malformés dans le but de déclencher des défaillances, et d’autre part, un détecteur conçu pour identifier et reconnaître ces attaques. En plus d'entraîner un outil de détection d’anomalies réseau. Cette approche offre aussi la possibilité de développer un générateur d’attaques artificielles qui pourraient être utilisées dans des datasets.

\section{Conclusion}
L'évolution de la 5G nécessite de changer nos méthodes de détection d'anomalies. Nous avons identifié un manque de corrélation entre les différents niveau d'analyse dans les travaux de l'état de l'art et proposons donc un système de détection multidomaine basé sur l'analyse sémantique, séquentielle et temporelle. Nous proposons donc un système explicable par design et facilitiant l'analyse humaine qui reste indispensable lors de l'utilisation d'un outil de détection d'anomalies. Le prochain objectif est de valider expérimentalement notre approche pour en démontrer la pertinence.

\section*{Remerciements}
Nous tenons à remercier le programme PEPR ``Réseaux du Futur`` de l'Agence Nationale de la Recherche (ANR) pour le soutien apporté à nos travaux dans le cadre du plan d'investissement France 2030. Nous remercions, en particulier, le projet NF-HiSec, financé sous la référence ANR-22-PEFT-0009, qui a joué un rôle essentiel dans la réalisation de ces travaux.

\bibliography{references}

\vspace{12pt}

\end{document}